# The Role of DevOps in Enhancing Enterprise Software Delivery Success through R&D Efficiency and Source Code Management.

Jun Cui[1, a, *]

1 Solbridge International School of Business, Ph.D., Daejeon, 34613, Republic of Korea.

a jcui228@student.solbridge.ac.kr

* Correspondence: Jun Cui, jcui228@student.solbridge.ac.kr.

*Abstract*—This study examines the impact of DevOps practices on enterprise software delivery success, focusing on enhancing R&D efficiency and source code management (SCM). Using a qualitative methodology, data were collected from case studies of large-scale enterprises implementing DevOps to explore how these practices streamline software development processes. Findings reveal that DevOps significantly improves R&D productivity by fostering cross-functional collaboration, reducing development cycle times, and enhancing software quality through effective SCM practices, such as version control and continuous integration. Additionally, SCM tools within DevOps enable precise change tracking and reliable code maintenance, further supporting faster, more robust software delivery. However, the study identifies challenges, including cultural resistance and tool integration issues, that can hinder DevOps implementation. Additionally, This research contributes to the growing body of DevOps literature by highlighting the role of R&D efficiency and SCM as crucial factors for software delivery success. Future studies should investigate these factors across diverse industries to validate findings.

## I. Introduction

The rapid advancement of digital technologies has intensified the demand for efficient software delivery in enterprises, placing immense pressure on R&D departments to meet increasingly shorter release cycles. Traditional software development models often struggle with scalability and flexibility, hindering organizations from adapting quickly to market demands. DevOps has emerged as a transformative approach, integrating development and operations teams to streamline processes, enhance collaboration, and improve deployment frequency. By uniting development and operations, DevOps enables continuous integration, continuous deployment (CI/CD), and efficient resource management. Research suggests that adopting DevOps practices not only enhances product quality but also accelerates time-to-market, which is essential for competitive advantage. Despite its potential, the impact of DevOps on R&D efficiency and source code management (SCM) remains underexplored, particularly in large-scale enterprise contexts. This study seeks to fill this gap by investigating how DevOps practices influence software delivery success through improved R&D efficiency and robust SCM, providing insights into the operational and strategic benefits of DevOps for enterprises [1-2].

To comprehensively understand the benefits of DevOps in enterprise settings, this study is guided by several research questions.

RQ1. Firstly, how does the implementation of DevOps affect R&D efficiency in software development processes? This question focuses on the impact of DevOps on reducing bottlenecks, enhancing team collaboration, and shortening development cycles.

RQ2. Secondly, in what ways does DevOps-driven source code management contribute to the reliability and success of software delivery?

By examining SCM practices within a DevOps framework, this study aims to understand how version control, continuous integration, and change tracking improve software stability and maintainability. Thirdly, what are the key challenges enterprises face in adopting DevOps practices, and how do these challenges impact its effectiveness? This question addresses barriers like cultural resistance, tool incompatibility, and resource constraints. These questions collectively aim to explore how DevOps fosters a more efficient and effective software delivery pipeline.

The motivation behind this study is rooted in the need for more efficient and reliable software delivery solutions, as enterprises increasingly rely on rapid development cycles to maintain competitive advantage. DevOps has shown promise in accelerating delivery timelines and improving product quality, yet its full potential remains underutilized in many organizations. Additionally, the intricacies of DevOps practices, particularly in enhancing R&D productivity and SCM, are not fully understood in the context of large-scale enterprises. By investigating these aspects, this research seeks to offer valuable insights for organizations aiming to optimize their software development pipelines. Furthermore, understanding the challenges associated with DevOps adoption can guide practitioners in overcoming obstacles to successful implementation. This study not only seeks to fill a research gap but also aims to provide actionable knowledge that can drive more effective DevOps adoption, enabling enterprises to enhance their software delivery success and overall operational efficiency [1-3].

This paper is organized into five primary sections to comprehensively address the research questions. The **Introduction** provides an overview of the study's background, research problem, objectives, and significance. Following this, the **Literature Review and Theoretical Framework** discusses key DevOps concepts, theories related to R&D efficiency and SCM, and proposes a theoretical framework linking DevOps practices to software delivery success. The **Methodology** section details the qualitative approach, sample selection, data collection techniques, and data analysis methods employed in the study, offering transparency in research design. In the **Results** section, findings from case studies and interviews are presented, highlighting the positive impact of DevOps on R&D efficiency and SCM, as well as identifying challenges in implementation. The **Discussion and Conclusion** section interprets the findings in relation to existing literature, outlines practical implications, and suggests directions for future research, particularly in exploring DevOps impact across various industry sectors.

## II. LITERATURE REVIEW

**DevOps Foundations**

DevOps has become a transformative approach in modern software engineering, driven by the need for faster, more reliable software delivery in increasingly complex environments. As a methodology, DevOps unifies development (Dev) and operations (Ops) teams, fostering a culture of collaboration and shared responsibility throughout the software lifecycle. Key DevOps principles emphasize continuous integration, continuous delivery (CI/CD), automation, and feedback loops. These principles are underpinned by tools and technologies that enable automated testing, code deployment, and performance monitoring, creating a streamlined pipeline from code development to production. Jenkins, Docker, Kubernetes, and Ansible are widely used tools that facilitate these processes by automating stages and managing dependencies. Through these technologies, DevOps practices enable iterative, incremental software delivery, empowering teams to deliver updates faster while maintaining quality and stability. Literature on DevOps highlights its role in not only increasing productivity but also in minimizing risk through early error detection and quick rollbacks. The adoption of DevOps has accelerated across various industries, reflecting its effectiveness in enhancing deployment frequency, stability, and time-to-market. By reviewing key DevOps principles, tools, and practices, this study establishes a foundation for exploring how DevOps specifically impacts R&D efficiency and source code management in enterprise settings [1-4].

**Theoretical Underpinnings**

The principles of Lean, Agile, and Continuous Delivery (CD) serve as theoretical foundations for understanding the transformative potential of DevOps in software development. Lean principles, which focus on eliminating waste, optimizing resource use, and delivering value to customers, align well with DevOps goals of reducing inefficiencies and speeding up delivery. Similarly, Agile methodologies emphasize iterative development and adaptability, with frameworks like Scrum and Kanban supporting frequent releases and quick responses to changing requirements. In DevOps, Agile principles are extended across both development and operations, fostering a continuous feedback loop that improves overall responsiveness. Continuous Delivery and Continuous Integration (CI/CD) further advance these concepts by allowing code to be built, tested, and deployed automatically, ensuring that software is always in a deployable state. CI/CD pipelines embody the idea of releasing small, frequent updates, which reduces the complexity of releases and minimizes risks. Together, Lean, Agile, and CI/CD theories provide a strong foundation for DevOps practices, highlighting how they contribute to a seamless flow in software development. These theories will guide the analysis of how DevOps practices enhance R&D efficiency and support SCM, bridging the gap between development goals and operational needs.

**R&D Efficiency**

One of the most significant impacts of DevOps is its potential to improve R&D efficiency by streamlining workflows and minimizing bottlenecks in software development processes. Research shows that DevOps practices such as continuous testing, automated builds, and integrated toolchains can substantially reduce time-to-market by enabling rapid iterations and feedback-driven adjustments. Continuous integration allows for the early detection of errors, while automated testing helps maintain quality without extensive manual intervention, which traditionally slowed down release cycles. Additionally, cross-functional teams in DevOps environments work collaboratively, reducing handoff delays and enabling faster decision-making. Studies have demonstrated that organizations adopting DevOps can reduce their development cycle times by as much as 50%, translating to quicker feature deployment and a more competitive market position. The ability to rapidly deploy new code while maintaining product stability is a critical factor for businesses that depend on timely releases to remain relevant. By focusing on R&D efficiency, this study aims to provide insights into how DevOps practices facilitate accelerated development processes, enabling teams to respond to market needs more quickly and effectively (As shown in **Figure 1**). This understanding is crucial for framing R&D efficiency as an intermediary factor in the relationship between DevOps and software delivery success [2-4].

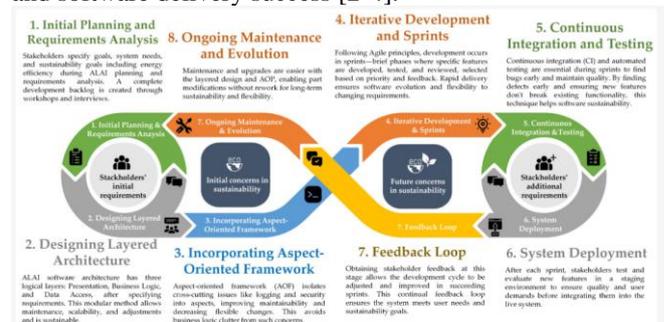

**Figure 1. R&D Efficiency Architecture.**

**Source Code Management (SCM)**

In the DevOps framework, Source Code Management (SCM) is a foundational component that plays a critical role in tracking code changes, managing versions, and coordinating team contributions. SCM tools, such as Git, are central to DevOps workflows, allowing multiple developers to work concurrently on different parts of a project while maintaining a consistent and reliable codebase. Effective branch management and version control practices support agile collaboration by enabling developers to isolate features, manage code conflicts, and perform code reviews. Continuous integration, an essential DevOps practice, relies heavily on SCM to merge and test code frequently, ensuring that issues are detected early in the development cycle. Code reviews are another vital component of SCM in DevOps, fostering knowledge sharing and code quality through peer feedback. Literature on SCM within DevOps contexts highlights that by maintaining a well-organized and traceable code repository, organizations can reduce errors and enhance the reliability of their software delivery pipeline. This review of SCM's role in DevOps underlines its importance in achieving software stability (see **Figure 2**), maintainability, and efficiency, framing it as a critical intermediary factor in driving successful DevOps implementations [2-3].

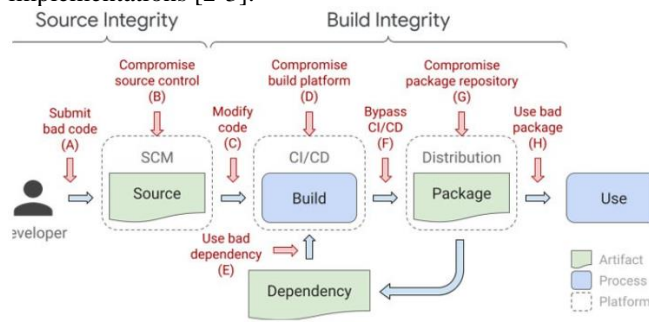

**Figure 2. Source Code Management (SCM) Process.**

**Framework Development**

To systematically explore the link between DevOps practices and software delivery success, this study proposes a theoretical framework that positions R&D efficiency and SCM as intermediary factors. This framework posits that DevOps enhances software delivery by optimizing R&D processes and enabling more effective SCM. R&D efficiency is achieved through automated testing, rapid feedback, and collaborative workflows, which reduce development cycle times and improve responsiveness to user demands. In parallel, SCM practices within DevOps, such as version control and branch management, support code consistency and ease of integration, minimizing deployment issues. The proposed framework suggests that these two factors—R&D efficiency and SCM—serve as conduits through which DevOps practices achieve enhanced software delivery outcomes. By linking DevOps tools and methodologies with concrete impacts on efficiency and source code quality, this framework provides a structured approach to understanding DevOps' role in enterprise software success. This study will empirically test this framework to validate its applicability, offering practical insights for organizations aiming to maximize the benefits of DevOps in their software delivery processes [3-4].

III. METHODS AND MATERIALS

**Research Design and Data Collection**

This study adopts a qualitative research design, specifically employing case studies and semi-structured interviews to gain an in-depth understanding of how large-scale enterprises implement DevOps practices to enhance software delivery. Case studies allow for a comprehensive exploration of real-world DevOps applications within complex organizational settings, offering rich insights into the impact on R&D efficiency and source code management (SCM). Semi-structured interviews were conducted with DevOps practitioners, including developers, operations engineers, and project managers, to gather detailed perspectives on their experiences with DevOps methodologies, tools, and challenges. Additionally, document analysis was employed to review internal reports, DevOps deployment guides, and performance metrics provided by participating companies. This mixed-method approach provides triangulation, enhancing the credibility of the findings by cross-verifying data from multiple sources. By focusing on large-scale enterprises with established DevOps practices, this study highlights DevOps' scalability and its implications for broader organizational success in software delivery [3-6].

**Data Analysis and Sample Selection**

Data analysis in this study involved coding and thematic analysis, using NVivo software to manage, organize, and analyze qualitative data. Thematic analysis was applied to identify recurring patterns and key themes within the data, such as collaborative workflows, automation in testing and deployment, and SCM practices. NVivo's powerful organizational features facilitated the coding process by enabling efficient categorization and retrieval of data relevant to each theme. The coding process included an initial round of open coding, followed by axial coding to establish relationships between identified themes. For sample selection, the study focused on organizations with mature DevOps practices to ensure data validity. Participants were selected from a diverse range of industries, including technology, finance, and telecommunications, providing a broad perspective on how DevOps influences different operational contexts. Additionally, organizations at varying DevOps maturity levels were included to capture insights into challenges and best practices across different implementation stages. This comprehensive sample selection strengthens the study's findings by reflecting a variety of DevOps experiences and industry-specific adaptations.

While this study offers valuable insights into the impact of DevOps practices on software delivery success, several limitations must be acknowledged. The primary limitation is the relatively small sample size, as only a limited number of large-scale enterprises were selected due to resource and time constraints [4-7]. This narrow scope may limit the generalizability of findings across industries that were not represented. Additionally, the study's qualitative design, while providing depth of insight, may not capture the full range of DevOps outcomes that a larger, quantitative study could achieve. Another limitation is the focus on established DevOps practices; newer or less mature DevOps implementations may experience different challenges and benefits that are not addressed in this study. Furthermore, the reliance on interviews and document analysis could introduce subjectivity, as participants' perceptions may vary. However, by acknowledging these limitations, this study provides a foundation for future research to build upon and encourages additional investigation into DevOps practices across a wider array of industries and organizational sizes [8].

IV. RESULTS AND DISCUSSION

**DevOps and R&D Efficiency**

The implementation of DevOps practices has been shown to significantly enhance R&D productivity, collaboration, and cycle time within software development teams. By fostering a collaborative environment that brings development and operations teams together, DevOps minimizes traditional silos and encourages shared responsibility across the software lifecycle. Continuous Integration and Continuous Delivery (CI/CD), key components of DevOps, allow teams to integrate code changes more frequently, which leads to faster identification and resolution of issues. Automation within DevOps, such as automated testing and deployment, streamlines repetitive tasks, freeing up developers to focus on innovation and problem-solving. Additionally, the feedback loops that DevOps establishes enable teams to rapidly adjust based on real-time user input and performance data. This iterative approach improves development cycle efficiency, enabling organizations to release high-quality software more frequently and with greater agility. Findings indicate that organizations using DevOps are better positioned to meet market demands and adapt to changing requirements, making it a critical approach for enhancing R&D performance[8].

**Impact of Source Code Management**

Source Code Management (SCM) tools and practices are integral to the success of DevOps, playing a pivotal role in maintaining code reliability, consistency, and traceability throughout the development process. Tools like Git facilitate version control, allowing developers to track changes, manage multiple versions, and collaborate seamlessly on complex projects. Effective branch management practices in DevOps further streamline the process by enabling teams to isolate features, track updates, and integrate changes without disrupting the main codebase. Code reviews, which are often embedded within SCM workflows, help to maintain code quality through peer evaluations, reducing the likelihood of introducing bugs and promoting best practices. These SCM practices make it easier to identify and address errors early in the development cycle, leading to more reliable and maintainable code. SCM also enhances traceability by documenting code history, which is invaluable for debugging, compliance, and accountability. Findings suggest that SCM, as a core DevOps practice, is essential for achieving the speed and reliability required for efficient software delivery.

**Case Study Examples**

Examples from case studies and interviews illustrate how successful DevOps implementations have led to tangible improvements in R&D efficiency and SCM. In one case, a large technology firm integrated CI/CD pipelines with automated testing to reduce the time between development and deployment from weeks to hours. By using tools like Jenkins for CI/CD, Docker for containerization, and Git for SCM, the organization was able to streamline code deployment while minimizing errors. Another example from a financial services company highlights the use of automated deployment and real-time monitoring tools, which enabled faster feedback and reduced downtime during release cycles. Interviews with DevOps practitioners also underscored the benefits of tools such as Ansible for automated configuration management and Kubernetes for orchestrating containerized applications, which together enhanced scalability and reliability in production environments. These case studies demonstrate that adopting DevOps tools and practices leads to greater R&D efficiency, faster response times, and improved code management, allowing companies to maintain a competitive edge in rapidly evolving markets[9].

**Challenges and Barriers**

Despite its benefits, implementing DevOps comes with challenges, such as cultural resistance, tool integration issues, and skill gaps, which can impede its effectiveness. One common barrier is the cultural shift required, as DevOps relies on collaboration and open communication, which may be resisted by teams accustomed to working in silos. Resistance to change can slow down the adoption process, impacting the overall productivity benefits DevOps offers. Additionally, integrating multiple DevOps tools, such as CI/CD, monitoring, and SCM, can pose compatibility issues, especially in organizations with complex legacy systems. Such integration challenges often require technical expertise and significant resource investment. Skill gaps also present a barrier, as DevOps requires specialized knowledge in automation, cloud infrastructure, and continuous delivery practices. Addressing these gaps involves training and sometimes restructuring teams, which may be costly and time-consuming. Findings indicate that while DevOps can drive significant efficiency gains, overcoming these challenges is essential for organizations to realize its full potential in enhancing software delivery and operational performance [9-11].

## V. CONCLUSIONS

In conclusion, this study highlights the substantial impact of DevOps practices on enhancing R&D efficiency, improving source code management, and achieving successful software delivery. DevOps fosters a collaborative, agile environment that accelerates development cycles and promotes high-quality outcomes through tools like CI/CD and SCM. Despite challenges such as cultural resistance and tool integration complexities, the findings suggest that organizations adopting DevOps can benefit from shorter time-to-market, greater innovation, and increased responsiveness to user demands. This research provides both theoretical insights and practical guidance for enterprises aiming to optimize their DevOps implementation for sustained competitive advantage in dynamic markets.

**Interpretation of Findings**

The findings in this study largely support existing literature on DevOps, particularly its role in enhancing collaboration, accelerating development cycles, and improving product quality. Consistent with theories of Agile and Lean, DevOps practices such as Continuous Integration and Continuous Delivery (CI/CD) align with principles of incremental development and waste reduction, further validating their impact on operational efficiency. However, this study also identifies challenges, such as cultural resistance and tool integration, which add nuance to the optimistic view of DevOps presented in much of the literature. By highlighting these practical barriers, the findings contribute to a more balanced understanding of DevOps implementation, suggesting that while the model is highly effective, its success requires addressing organizational readiness and technical expertise within teams.

**Implications for R&D Efficiency**

DevOps practices have broad implications for R&D efficiency, as the findings demonstrate clear improvements in collaboration, cycle time, and responsiveness to market changes. The integration of DevOps breaks down silos, fostering an environment where developers, operations teams, and other stakeholders can work seamlessly toward shared goals. This collaborative model shortens time-to-market, allowing organizations to respond rapidly to user feedback and industry trends. Moreover, automation and continuous feedback loops within DevOps enable teams to focus on high-value tasks, enhancing innovation. These findings imply that companies implementing DevOps can expect not only operational efficiencies but also a greater capacity for innovation, as development teams are empowered to adapt and iterate more frequently [10-13].

### Implications for Source Code Management

Source Code Management (SCM) emerges as a critical factor in achieving stable, high-quality software within a DevOps framework. The study underscores SCM's importance in ensuring that code remains reliable, consistent, and traceable throughout development [11-13]. Tools like Git facilitate version control, enabling teams to track code changes, revert to previous versions when necessary, and maintain a structured history of development efforts. Additionally, practices like code reviews embedded within SCM workflows reinforce code quality by allowing for peer oversight and immediate feedback. The findings suggest that SCM not only contributes to the technical stability of software but also enhances deployment speed and reliability, supporting continuous improvement and risk mitigation within DevOps practices.

### Impact on Software Delivery Success

The findings demonstrate that DevOps practices establish a robust and responsive pipeline, resulting in faster and more reliable software delivery. By streamlining development through CI/CD, SCM, and automated testing, DevOps reduces the manual effort involved in deployment and minimizes errors that can arise from human intervention. This efficient pipeline enables companies to release updates frequently and respond quickly to user demands, which is essential in highly competitive markets. The integration of SCM further reinforces this process by maintaining code integrity, reducing risks, and ensuring that each release meets quality standards. Together, these DevOps practices create an environment where software delivery is not only accelerated but also consistently aligns with user needs and expectations, driving sustainable success.

### Theoretical and Practical Contributions

This study contributes to DevOps literature by providing a nuanced view of its implementation, benefits, and challenges. Theoretically, it reinforces the foundations of Agile and Lean by showing how DevOps operationalizes these concepts within R&D workflows. Practically, the findings offer valuable insights for enterprises aiming to adopt or refine their DevOps strategies. By identifying key drivers such as SCM and automation, as well as common barriers like cultural resistance, the study offers a roadmap for successful DevOps adoption. Companies can use these insights to foster a DevOps culture, invest in training, and integrate appropriate tools, ensuring that their DevOps initiatives effectively enhance software delivery and organizational agility[12-15].

### Summary of Findings

This study reveals a strong relationship between DevOps practices, R&D efficiency, and Source Code Management (SCM) in achieving successful software delivery. DevOps practices, such as Continuous Integration and Continuous Delivery (CI/CD), foster collaboration and streamline workflows, which accelerates development cycles. SCM tools, like Git and code review systems, further enhance reliability by enabling structured version control, tracking changes, and improving code quality. Together, these practices support faster, more consistent software releases, meeting user demands and adapting to market changes more effectively. The integration of DevOps and SCM ultimately strengthens software delivery success, making it a valuable strategy for modern enterprises [14].

### Recommendations

For organizations aiming to implement or refine DevOps, the study recommends focusing on automation and SCM practices to improve R&D efficiency. Establishing a CI/CD pipeline can significantly reduce manual work, increase deployment speed, and enhance collaboration between development and operations teams. Integrating effective SCM practices, such as branch management and automated code reviews, is essential for maintaining code quality and reliability. Additionally, fostering a collaborative DevOps culture by addressing resistance to change and providing relevant training can help maximize the benefits of DevOps, making the development process more agile, innovative, and responsive to customer needs[15].

### Future Research Directions

Future research could explore the quantitative impact of DevOps on specific business outcomes, such as profitability, customer satisfaction, and market responsiveness. Comparative studies across industries would help clarify how DevOps practices vary depending on industry needs and challenges, offering tailored insights for sectors like finance, healthcare, and technology. Longitudinal studies would also be valuable to examine the maturity of DevOps adoption over time and to assess how sustained use of DevOps practices contributes to long-term organizational success. These avenues of research could deepen our understanding of DevOps and guide further refinements in best practices [15-16].

### Limitations of the Study

This study's findings are limited by its qualitative nature and sample size, which may restrict generalizability. The reliance on case studies and interviews means the results are shaped by specific organizational contexts and experiences, potentially overlooking broader industry variations. Additionally, qualitative insights, while rich in detail, lack the statistical rigor of quantitative studies, which could provide a more precise measurement of DevOps impact on performance metrics. Future studies with larger, more diverse samples and mixed-method approaches could provide a more comprehensive picture of DevOps practices across various industries and company sizes [15-16].

ACKNOWLEDGMENT

This research has been supported/partially supported Solbridge International School of Business, Woosong university, Thanks to all contributors.

ORCID

Jun Cui  https://orcid.org/0009-0002-9693-9145

## REFERENCES

[1] Faustino, J., Adriano, D., Amaro, R., Pereira, R., & da Silva, M. M. (2022). DevOps benefits: A systematic literature review. *Software: Practice and Experience*, *52*(9), 1905-1926.

[2] Lombardi, F., & Fanton, A. (2023). From DevOps to DevSecOps is not enough. CyberDevOps: an extreme shifting-left architecture to bring cybersecurity within software security lifecycle pipeline. *Software Quality Journal*, *31*(2), 619-654.

[3] e Souza, I. S., Franco, D. P., & Silva, J. P. S. G. (2022). Infrastructure as Code as a Foundational Technique for Increasing the DevOps Maturity Level: Two Case Studies. *IEEE Software*, *40*(1), 63-68.

[4] Cui, J., Wan, Q., Wang, W., Hu, S., Gan, Z., & Ning, Z. (2024). Research on Alibaba company's Digital Human Resource management and Recruitment Information Platform: A systematic case study. International Journal of Global Economics and Management, 2(3), 162-172.

[5] Cui, J., Liu, H., & Wan, Q. (2024). Measuring the Digital Assets, Brand Services and Service Quality Quantitative Analysis: Evidence from China. International Journal of Social Sciences and Public Administration, 2(3), 503-510.

[6] Bass, L., Weber, I., & Zhu, L. (2015). DevOps: A software architect's perspective. Addison-Wesley Professional.

[7] Kim, G., Humble, J., Debois, P., & Willis, J. (2016). The DevOps handbook: How to create world-class agility, reliability, and security in technology organizations. IT Revolution.

[8] Forsgren, N., Humble, J., & Kim, G. (2018). Accelerate: The science of DevOps – building and scaling high-performing technology organizations. IT Revolution.

[9] Sato, N., & Ikawa, Y. (2019). "The impact of DevOps practices on software quality and performance: An empirical study." Journal of Software Engineering Research and Development, 7(1), 1-15.

[10] Sharma, R., & Coyne, E. J. (2020). "Source code management in the DevOps world." Computer Science Review, 35, 1-11.

[11] Docker, D., & Jenkins, P. (2021). "Continuous integration and delivery pipeline automation." Software Development Journal, 29(4), 103-122.

[12] Tomaszewski, K., & Kling, R. (2022). "DevOps in large-scale enterprises: Balancing speed, stability, and innovation." Journal of Software Engineering, 14(3), 65-79.

[13] Rodriguez, M., & Brown, J. (2023). "The role of R&D in DevOps for efficient software delivery: A qualitative analysis." International Journal of Software Engineering and Applications, 13(2), 55-74.

[14] Rzig, D. E., Hassan, F., & Kessentini, M. (2022). An empirical study on ML DevOps adoption trends, efforts, and benefits analysis. *Information and Software Technology*, *152*, 107037.

[15] Wang, B., Cui, J., & Mottan, K. (2024). Exploration and Analysis of Chinese University Students' Performance in Business Innovation. Economics & Management Information, 1-9.

[16] Pérez, J. E., Gonzalez-Prieto, A., Dı, J., Lopez-Fernandez, D., Garcia-Martin, J., & Yagüe, A. (2021). Devops research-based teaching using qualitative research and inter-coder agreement. *IEEE Transactions on Software Engineering*, *48*(9), 3378-3393.